\documentclass[a4paper,12pt]{article}

\usepackage{epsfig,graphicx}
\usepackage{epstopdf}
\usepackage[font=footnotesize, labelfont=bf]{caption,subcaption}
\usepackage{booktabs}
\usepackage{dcolumn}
\usepackage{makecell}
\usepackage{multirow}
\usepackage{amstext}

\begin{document}

\begin{center}
{\large\bf Decay behaviors of the $P_c$ hadronic molecules}

{C.W. Shen$^{1,2,\P}$}, {Y.H. Lin$^{1,2}$}

$^1${Key Laboratory of Theoretical Physics, Institute of Theoretical Physics,
Chinese Academy of Sciences, Beijing 100190,China}

$^2${University of Chinese Academy of Sciences (UCAS), Beijing 100049, China}

$^\P${E-mail: shencw@itp.ac.cn}
\end{center}

\centerline{\bf Abstract}
In this proceeding, we present our recent work on decay behaviors of the $P_c$ hadronic molecules,
which can help to disentangle the nature of the two $P_c$ pentaquark-like structures.
The results turn out that the relative ratio of the decays of $P^+_c(4380)$ to $\bar{D}^* \Lambda_c$ and $J/\psi p$
is very different for $P_c$ being a $\bar D^*\Sigma_c$ or $\bar D\Sigma_c^*$ bound state with $J^P=\frac{3}{2}^-$.
And from the total decay width, we find that $P_c(4380)$ being a $\bar D\Sigma_c^*$ molecule state with $J^P=\frac{3}{2}^-$
and $P_c(4450)$ being a $\bar D^*\Sigma_c$ molecule state with $J^P=\frac{5}{2}^+$ is more favorable to the experimental data.

\section{Introduction}

In 2015, LHCb Collaboration observed two hidden charm pentaquark-like states
in the $J/\psi p$ invariant mass distribution in $\Lambda^0_b \to J/\psi K^- p$ decays~\cite{Aaij:2015tga}.
The significance of them are both larger than 9 standard deviations.
According to LHCb Collaboration's fit results, the masses and widths of the $P^+_c(4380)$ and $P^+_c(4450)$
are $(4380\pm8\pm29)\ \mathrm{MeV}$, $(205\pm 18\pm 86)\ \mathrm{MeV}$
and $(4449.8 \pm 1.7\pm 2.5) \ \mathrm{MeV}$, $(39\pm 5\pm 19)\ \mathrm{MeV}$, respectively.
It also shows that the preferred spins of these two states are one having spin 3/2 and the other 5/2 with opposite parities.

Various theoretical schemes are proposed to explore these two $P_c$ states' inner structures.
In Ref.~\cite{Maiani:2015vwa,Anisovich:2017aqa}, the two $P_c$ states are discussed as the diquark-diquark-antiqurk states in the $\bar{c}[cu][ud]$ configuration.
The diquark-triquark system is used to study them in Ref.~\cite{Lebed:2015tna,Zhu:2015bba}.
The kinematic effect is considered in Ref.~\cite{Guo:2015umn,Liu:2015fea} to explain the narrow width of $P_c(4450)$ state.
Also many groups conjecture the two $P_c$ states being $\bar D^{(*)} \Sigma_c^{(*)}$ hadronic molecules~\cite{Chen:2015loa,Roca:2015dva,He:2015cea,Lu:2016nnt},
and this scheme has already been used to predict these pentaquarklike molecular states before the discovery of the two $P_c$ states in Ref.~\cite{Wu:2010vk,Wu:2010jy}.
Since the mass thresholds of the $\bar D\Sigma_c^*$ and $\bar D^*\Sigma_c$ are 4387 MeV and 4461 MeV, respectively,
which locate quite close to the mass region of these two $P_c$ states, we interpret naturally them as the hadronic
molecular states composed of either $\bar D\Sigma_c^*$ or $\bar D^*\Sigma_c$.

In this work we will estimate firstly the partial decay widths of the $P_c(4380)$ into
the $\bar D \Lambda^+_c$ and $J/\psi p$ channels assuming it is an S-wave $\bar D\Sigma_c^*$ or $\bar D^*\Sigma_c$ hadronic molecular state.
We found that supposing $P_c(4380)$ is a $\bar D^*\Sigma_c$ or $\bar D\Sigma_c^*$ hadronic molecular state with $J^P=\frac{3}{2}^-$,
the relative ratios of decaying to $\bar D \Lambda^+_c$ and $J/\psi p$ are very different.
This can be applied to distinguish the nature of $P_c(4380)$ and examined by experiments in the future.
It also can shed light on the large decay width of $P_c(4380)$.
Then we calculate the partial decay widths of the two $P_c$ states into various possible final states.
The decay patterns differ a lot for the $P_c$ states having different components with different spin-parity, and the total decay widths are also different greatly. Based on these decay patterns, we also estimate the production of two $P_c$ states in photo- and pion- induced reactions, which shows large cross sections for the $\bar{D^*}\Lambda_c$ and $\bar{D}\Lambda_c$ production through the s-channel exchange of $P_c$ states.

\section{Framework}

\begin{figure}[htbp]
 \includegraphics[width=1.0\textwidth]{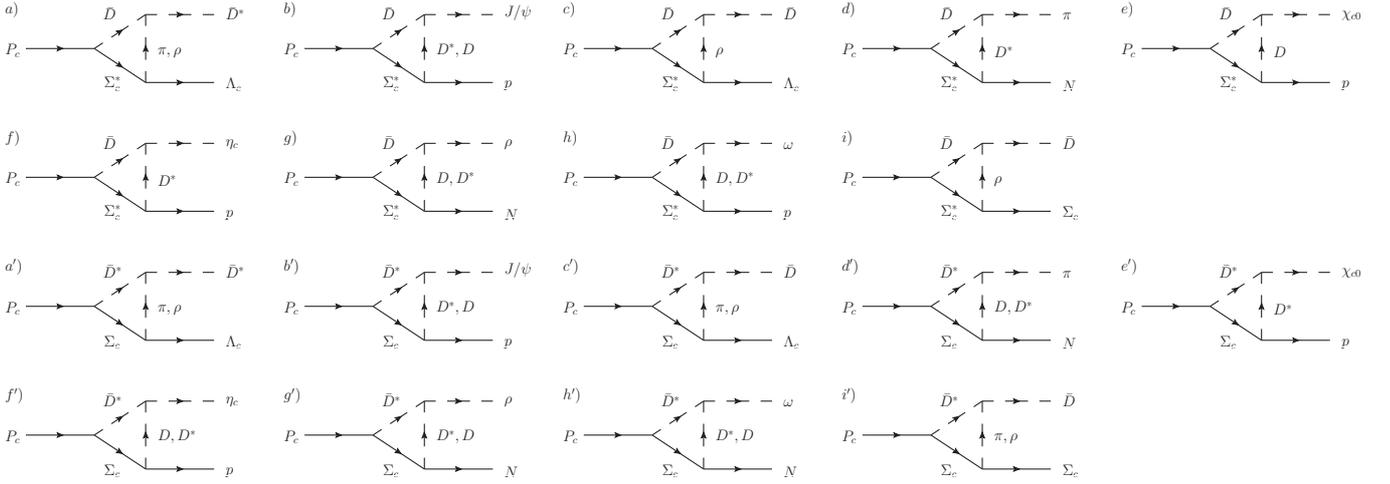}
 \centering
 \caption{The decays of the $P_c$ state via meson exchange as a (I): $\bar{D} \Sigma_c^*$ molecule.
$a$) $\bar D^*\Lambda_c$ channel.
$b$) $J/\psi p$ channel.
$c$) $\bar D\Lambda_c$ channel.
$d$) $\pi N$ channel.
$e$) $\chi_{c0}p$ channel.
$f$) $\eta_c p$ channel.
$g$) $\rho N$ channel.
$h$) $\omega p$ channel.
$i$) $\bar D\Sigma_c$ channel.
(II): a $\bar{D}^* \Sigma_c$ molecule.
$a^{\prime}$) $\bar D^*\Lambda_c$ channel.
$b^{\prime}$) $J/\psi p$ channel.
$c^{\prime}$) $\bar D\Lambda_c$ channel.
$d^{\prime}$) $\pi N$ channel.
$e^{\prime}$) $\chi_{c0}p$ channel.
$f^{\prime}$) $\eta_c p$ channel.
$g^{\prime}$) $\rho N$ channel.
$h^{\prime}$) $\omega p$ channel.
$i^{\prime}$) $\bar D\Sigma_c$ channel.}
 \label{feynman}
\end{figure}

Firstly, we calculate the partial decay widths of $P_c(4380)$ into $\bar D \Lambda^+_c$ and $J/\psi p$ final states by treating $P_c(4380)$ as an S-wave hadronic molecular state of either $\bar D\Sigma_c^*(2520)$ or $\bar D^*\Sigma_c(2455)$ with $J^P=\frac{3}{2}^-$.
These two decays are proceeded through triangular diagrams as shown in $a$), $b$), $a^{\prime}$) and $b^{\prime}$) of Fig.\ref{feynman}.
The calculation is accomplished in the framework of effective Lagrangian approach.
The detailed Lagrangians for different types of vertices and the involved coupling constants can be found in Ref.~\cite{Shen:2016tzq,Lin:2017mtz},
since they are not shown here for simplicity.

To make the four-dimension loop integrals in the amplitudes convergent, a Gaussian regulator is added in the first vertex with the form of Eq.(\ref{ff0}) after the time-part integration has been done with the residual theorem.
The value of cutoff $\Lambda_0$ is varied in the range of 0.5 $\mathrm{GeV}$ to 1.2 $\mathrm{GeV}$ and $\textbf{q}$ is the spatial part of the loop momentum $q$.

\begin{equation}
\Phi_{P_c}(\textbf{q}^2/\Lambda_0^2) \equiv \exp( - \textbf{q}^2/\Lambda_0^2)
\label{ff0}
\end{equation}

Then we study the decay properties of $P_c(4380)$ being a $\bar D\Sigma_c^*(2520)$ molecule or $\bar D^*\Sigma_c(2455)$ molecule with $J^P={\frac32}^-$
and $P_c(4450)$ being a $\bar D^*\Sigma_c(2455)$ molecule with $J^P={\frac32}^-$ or ${\frac52}^+$.
The possible final states we considered are listed in Table~\ref{channel}
and the corresponding feynman diagrams with pseudoscalar and vector mesons exchange included are shown in Fig.\ref{feynman}.
The effective Lagrangians and couplings can be also found in Ref.~\cite{Shen:2016tzq,Lin:2017mtz}.

\begin{table}[htbp]
\centering
\caption{
\label{channel}
All possible final states for the $P_c(4380)$ with $J^P={\frac32}^-$ and $P_c(4450)$ with $J^P={\frac32}^-$ or ${\frac52}^+$.}
\begin{tabular}{l|c}
\Xhline{1pt}
\thead{Initial state} & \thead{Final states} \\
\Xhline{0.8pt}
$P_c(4380)(\bar{D} \Sigma_c^*)$ 	 & $\bar D^*\Lambda_c$, $J/\psi p$, $\bar D\Lambda_c$, $\pi N$, $\chi_{c0}p$, $\eta_c p$, $\rho N$, $\omega p$, $\bar D\Sigma_c$ 	\\
$P_c(4380)(\bar{D}^* \Sigma_c)$ 	 & $\bar D^*\Lambda_c$, $J/\psi p$, $\bar D\Lambda_c$, $\pi N$, $\chi_{c0}p$, $\eta_c p$, $\rho N$, $\omega p$, $\bar D\Sigma_c$	\\
\Xhline{0.8pt}
$P_c(4450)(\bar{D}^* \Sigma_c)$ 	 &  $\bar D^*\Lambda_c$, $J/\psi p$, $\bar D\Lambda_c$, $\pi N$, $\chi_{c0}p$, $\eta_c p$, $\rho N$, $\omega p$, $\bar D\Sigma_c$, $\bar{D} \Sigma_c^*$      \\
\Xhline{1pt}
\end{tabular}
\end{table}

It should be mentioned that the coupling constants which describe the strength of the interaction between the composite state and its compositions in the first vertex can be estimated with the compositeness condition.
Since the corresponding interaction for the $J^P={\frac52}^+$ $P_c(4450)$ state is P-wave, the coupling in this vertex will depend on some parameters which are included necessarily to make the loop integration convergent. In our case, we use the cutoff $\Lambda_0$ same as the form factor Eq.(\ref{ff0}) for simplicity.

Another off-shell form factor is introduced in our calculation for the exchanged meson with the form of Eq.(\ref{ff1}),
where $m$ and $q$ are the mass and momentum of the exchanged meson.
And the off-shell form factor's cutoff $\Lambda_1$ varies from 1.5 $\mathrm{GeV}$ to 2.4 $\mathrm{GeV}$.

\begin{equation}
F(q^2) = \frac{\Lambda_1^4}{(m^2 - q^2)^2 + \Lambda_1^4}
\label{ff1}
\end{equation}

\section{Results and discussions}

The ratio $R$ for the $P_c(4380)$ state decaying to $\bar D^* \Lambda_c$ channel and $J/\psi p$ channel is defined in Eq.(\ref{ratio}).
Here we only estimate the order of the ratio results, since there are many variable factors cannot be determined.
For $P_c(4380)$ being a $\bar{D} \Sigma_c^*$ hadronic molecule,
the partial decay width of $\bar{D}^* \Lambda_c$ channel is at least one order of magnitude larger than that of $J/\psi p$ channel.
However, the partial decay width of these two channels are in the same order if $P_c(4380)$ is assumed as a $\bar{D}^* \Sigma_c$ hadronic molecule.

\begin{eqnarray}
R_\text{I} = \frac{\Gamma(P_c(4380) \to \bar{D} \Sigma_c^* \to \bar{D}^*
\Lambda_c)} {\Gamma(P_c(4380) \to \bar{D} \Sigma_c^* \to J/\psi p)} \sim 10 \nonumber \\
R_\text{II} = \frac{\Gamma(P_c(4380) \to \bar{D}^* \Sigma_c \to \bar{D}^*
\Lambda_c)} {\Gamma(P_c(4380) \to \bar{D}^* \Sigma_c \to J/\psi p)} \sim 1
\label{ratio}
\end{eqnarray}

It is found that the partial width of $P_c(4380)$ being a $\bar{D} \Sigma_c^*$ hadronic molecule
should be much larger than being a $\bar{D}^* \Sigma_c$ hadronic molecular,
as there is a large contribution from $\bar{D}^* \Lambda_c$ channel in the former case.
This conclusion is also made by using the non-relativistic formalism with heavy quark spin symmetry taken into consideration,
whose details are shown in Appendix B of Ref.~\cite{Shen:2016tzq}.
It might explain the broad width of $P_c(4380)$ claimed by LHCb Collaboration.

Considering the possible channels in Table~\ref{channel},
we find that $\bar D^* \Lambda_c$, $J/\psi p$ and $\bar D \Lambda_c$ are the three dominant decay channels.
The dependance of total decay widths of the two $P_c$ states and branching fractions of these three channels
on cutoff $\Lambda_0$ and $\Lambda_1$ are shown in Ref.~\cite{Lin:2017mtz}.
Choosing $\Lambda_0=1.0\ \mathrm{GeV}$ and $\Lambda_1=2.0\ \mathrm{GeV}$,
the numerical results we obtained are listed in Table~\ref{total}.

\begin{table}[htbp]
\centering
\caption{\label{total}Partial widths of $P_c(4380)$ as a $\bar D \Sigma_c^*$ molecule
and a $\bar D^*\Sigma_c$ molecule with $J^P=\frac32^-$, and $P_c(4450)$ as a $\bar D^*\Sigma_c$ molecule with $J^P=\frac32^-$ and $\frac52^+$,
into different final states with $\Lambda_0=1.0\ \mathrm{GeV}$, $\Lambda_1=2.0\ \mathrm{GeV}$.
The short bars denote that this channel is closed or the corresponding contribution is negligible.}
\begin{tabular}{l|*{4}{c}}
\Xhline{1pt}
\multirow{3}*{Mode} & \multicolumn{4}{c}{Widths ($\mathrm{MeV}$)} \\
\Xcline{2-5}{0.4pt}
& \multicolumn{2}{c}{$P_c(4380)$} & \multicolumn{2}{c}{$P_c(4450)$} \\
\Xcline{2-3}{0.4pt}\Xcline{4-5}{0.4pt}
& \multicolumn{1}{c}{$\bar D \Sigma_c^*$(${\frac32}^-$)} & \multicolumn{1}{c}{$\bar D^*\Sigma_c$(${\frac32}^-$)} & \multicolumn{1}{c}{$\bar D^*\Sigma_c$(${\frac32}^-$)} & \multicolumn{1}{c}{$\bar D^*\Sigma_c$(${\frac52}^+$)} \\
\Xhline{0.8pt}
$\bar D^*\Lambda_c$ 	 & 110.4  	 & 28.6   	 & 63.8 	& 16.3\\
$J/\psi p$ 		 	     & 2.7    	 & 19.8  	 & 17.7     & 2.6\\
$\bar D\Lambda_c$  	     & 1.2  	 & 13.7  	 & 36.0     & 14.9\\
$\pi N$ 			 	 & 0.08  	 & 0.06      & 0.06     & 0.03\\
$\chi_{c0}p$ 		 	 & 0.8    	 & 0.002  	 & 0.01     & 0.001\\
$\eta_c p$ 		 	     & 0.2    	 & 0.05   	 & 0.1 	    & 0.02\\
$\rho N$ 			  	 & 1.6   	 & 0.4    	 & 0.2      & 0.1 \\
$\omega p$ 		 	     & 6.1  	 & 1.3   	 & 0.8      & 0.4\\
$\bar D\Sigma_c$ 	  	 & 0.01 	 & 0.09    	 & 1.1      & 0.2\\
$\bar D\Sigma^*_c$ 	  	 & -	     & -    	 & 8.9      & 0.5\\
$\bar D\Lambda_c \pi$ 	 & 7.5   	 & -    	 & -        & -\\
\Xhline{0.8pt}
Total 				 	 & 130.6  	 & 64.0   	 & 128.7    & 35.0\\
\Xhline{1pt}
\end{tabular}
\end{table}

Our results show that $P_c(4380)$ being a $\bar D \Sigma_c^*$ molecule will have larger width than being a $\bar D^*\Sigma_c$ molecule,
and the width of $P_c(4450)$ having spin-parity $\frac52^+$ is several times smaller than
having spin-parity $\frac32^-$ when supposing it is a $\bar D^*\Sigma_c$ molecule.
Compared with the widths reported by the LHCb Collaboration,
the $P_c(4380)$ in the $J^P=\frac32^-$ $\bar D \Sigma_c^*$ picture
and $P_c(4450)$ in the $J^P=\frac52^+$ $\bar D^* \Sigma_c$ picture in more favorable.

The cross sections of these reactions through s- and u- channel exchange of two $P_c$ states with $\gamma p$ and $\pi p$ incident and $\bar D^* \Lambda_c$, $J/\psi p$ and $\bar D \Lambda_c$ outgoing respectively are calculated in Ref.~\cite{Lin:2017mtz}.
The productions for $\bar D^* \Lambda_c$ and $\bar D \Lambda_c$ final states are much larger than the $J/\psi p$ final state.
It is helpful to figure out the nature of these two pentaquarklike $P_c$ states to seek for them in the $\bar D^* \Lambda_c$ and $\bar D \Lambda_c$ production in the future experiments.

\section*{Acknowledgment}
We thank Bing-Song Zou, Feng-Kun Guo and Ju-Jun Xie for useful suggestions and the effects on this work.
This project is supported by NSFC under Grant
No. 11261130311 (CRC110 cofunded by DFG and
NSFC) and Grant No. 11647601, and by the Thousand
Talents Plan for Young Professionals, and by the CAS Key
Research Program of Frontier Sciences under Grant
No. QYZDB-SSW-SYS013.

\end{document}